\documentclass[pra,twocolumn,showpacs]{revtex4-1}
\usepackage{graphicx}
\usepackage{color}
\newcommand{\bra}[1]{\left\langle{#1}\right\vert}
\newcommand{\ket}[1]{\left\vert{#1}\right\rangle}

\let\oldhat\hat
\renewcommand{\hat}[1]{\oldhat{\mathbf{#1}}}
\begin{document}
\title{Experimental construction of generic
three-qubit states and their reconstruction from two-party
reduced states on an NMR quantum information
processor}
\author{Shruti Dogra}
\email{shrutidogra@iisermohali.ac.in}
\affiliation{Department of Physical Sciences, Indian
Institute of Science Education \& 
Research (IISER) Mohali, Sector 81 Mohali, 
Manauli PO 140306 Punjab India.}
\author{Kavita Dorai}
\email{kavita@iisermohali.ac.in}
\affiliation{Department of Physical Sciences, Indian
Institute of Science Education \& 
Research (IISER) Mohali, Sector 81 Mohali, 
Manauli PO 140306 Punjab India.}
\author{Arvind}
\email{arvind@iisermohali.ac.in}
\affiliation{Department of Physical Sciences, Indian
Institute of Science Education \& 
Research (IISER) Mohali, Sector 81 Mohali, 
Manauli PO 140306 Punjab India.}
\begin{abstract}
We experimentally explore the state space of three
qubits on an NMR quantum information processor.
We construct a scheme to experimentally realize a
canonical form for general three-qubit states up
to single-qubit unitaries.  This form involves a
non-trivial combination of GHZ and W-type 
maximally entangled states of three qubits.  
The general circuit that we have constructed
for the generic state reduces to those for 
GHZ and W states as special cases.
The experimental construction of a generic
state is carried out for a nontrivial set of
parameters and the good fidelity of preparation
is confirmed by complete state tomography. The
GHZ and W-states are constructed as special
cases of the general experimental scheme.
Further, we experimentally demonstrate a curious
fact about three-qubit states, where for almost
all pure states, the two-qubit reduced states
can be used to reconstruct the full three-qubit
state. For the case of a generic state and for
the W-state, we demonstrate this method of reconstruction
by comparing it with the directly tomographed
three-qubit state. 
\end{abstract}
\pacs{03.67.Lx, 03.67.Bg}
\maketitle
\section{Introduction}
\label{intro}
While a qubit is considered to be a building block
for quantum information processing, the actual
quantum computer invariably involves complex
states of multiple qubits~\cite{nielsen-book-02}.
The transition from one to two qubits is of
fundamental importance because it is the two-qubit
system for which we can have entangled states and
hence a nontrivial quantum advantage for
information
processing~\cite{horodecki-rmp-2009,ladd-nature-2010}.
The manipulation of two-qubit states is
qualitatively more difficult than that for a
single qubit. As a matter of fact, the dynamics of
a single qubit finds a classical analog in
polarization optics~\cite{arvind-josab-2007}, and
it is only when we create entangled states of two
qubits, do the nontrivial quantum aspects
emerge~\cite{wootters-prl-98}. It may appear that
moving from two qubits to several qubits is merely
a matter of detail.  However, this is not the case
and new quantum aspects emerge for a three-qubit
system, which  is the simplest system for which
the concept of multi-partite entanglement can be
introduced.  Unlike the two-qubit case, the
maximally entangled states of three qubits are not
equivalent up to local unitary transformations and
instead fall into two inequivalent classes, namely
the GHZ and W classes of
states~\cite{chen-pra-06}.
In contradistinction to the two-qubit case, a canonical form for three
qubits turns out to be nontrivial and involves a combination
of GHZ and W states.   It has been shown that all pure
states of a system of three qubits are equivalent under
local unitary transformations to a canonical state  with
five independent non-zero
real parameters~\cite{acin-prl-00,carteret-jmp-00,acin-prl-01,vicente-prl-12,vertesi-pra-14}.
While one-qubit reduced states  have
information about the amount of entanglement in
a two-qubit pure state, they do not uniquely
determine the state.  On the other hand, it turns
out that almost every three-qubit pure state is
completely determined by its two-qubit reduced
density matrices and there is no more information
in the full quantum state than what is already
contained in the three possible two-qubit reduced
states~\cite{linden-prl-1-02,diosi-pra-04,cavalcanti-pra-05}.  
It is indeed
somewhat surprising that even when nontrivial
multi-partite entanglement is present, the
``parts'' can determine the  ``whole''.

There have been several experimental
implementations of tripartite-entangled W and GHZ
states using different physical
resources~\cite{laflamme-proc-98,nelson-pra-00,
mikami-prl-05,resch-prl-05,roos-science-04}.
GHZ and W states have been used as a resource in a
quantum prisoner's dilemma game~\cite{han-pla-02},
to simulate the violation of Bell-type
inequalities~\cite{ren-pla-09}, in quantum
erasers~\cite{teklemariam-chaos-03,teklemariam-pra-02}
and complementarity
measurements~\cite{peng-pra-08}, quantum key
distribution~\cite{kempe-pra-99}, quantum secret
sharing~\cite{hillery-pra-99} and quantum
teleportation~\cite{yeo-prl-06}.
In the context of NMR quantum computing, GHZ and W
states have been generated on a one-dimensional
Ising chain~\cite{rao-ijqi-12,gao-pra-13}, their
decoherence properties
studied~\cite{kawamura-ijqc-06}, and their ground
state phase transitions investigated in a system
with competing many-body
interactions~\cite{peng-prl-09,peng-pra-10}.

This work has two main results:~(a) We prescribe a
scheme to create generic states of three qubits
and implement it on an NMR quantum computer.  The
complete class of separable, biseparable  and
maximally entangled three-qubit states can be
generated using our scheme; (b) We experimentally
demonstrate the reconstruction of generic
three-qubit states from their two-qubit reduced
marginals.  The material in this paper is
organized as follows:~Section~\ref{implement}
describes the NMR implementation of a generic
state with a nontrivial five-parameter set, and
the implementations of the GHZ and W-states as
special cases of the general scheme.  The density
matrices of all the states are reconstructed by
using an optimal set of NMR state tomography
experiments.  Section~\ref{2tomo} describes the
three-qubit state reconstruction from their
two-party reduced states for a generic state and
for the W-state.  By comparing the state
tomographs obtained from the two-qubit marginals
and by a full tomography of the three-qubit state
we demonstrate that, reduced two-qubit density matrices are
indeed able to capture all information about the
full three-qubit state.  Section~\ref{concl}
contains some concluding remarks.
\section{NMR implementation}
\label{implement}
\begin{figure}[h]
\includegraphics[scale=1]{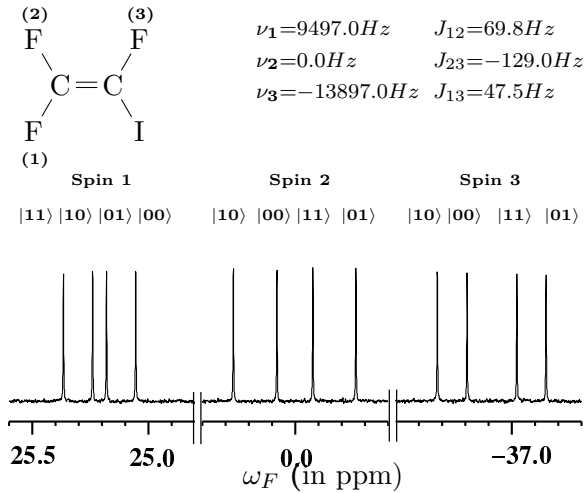}
\caption{
Molecular structure, NMR parameters and
${}^{19}$F thermal equilibrium spectrum of trifluoroiodoethylene.
The three fluorine spins in the
molecule
are marked as the corresponding qubits.
The table summarizes the relevant
NMR parameters i.e. resonance frequencies
$\nu_i$ and J-coupling constants.
The
${}^{19}$F spectrum is obtained after a $\pi/2$
readout pulse on the thermal equilibrium state.
The resonance lines of each qubit are labeled by
the corresponding logical states of the other
two qubits in the computational basis.
\label{system}
}
\end{figure}
The three-qubit system that we use for
NMR quantum information processing is 
the molecule trifluoroiodoethylene 
dissolved in deuterated acetone.
The three qubits were encoded using the
${}^{19}$F nuclei.
The Hamiltonian of the three-qubit system in the
rotating frame is given by
\begin{equation}
H = \sum_{i=1}^{3} \nu_i I_{iz} +  \sum_{i <
j, i=1}^{3} J_{ij} I_{iz}
I_{jz}
\end{equation}
where $I_{iz}$ is the single-spin Pauli 
angular momentum operator, $\nu_i$ are the Larmor frequencies of the
spins and $J_{ij}$ are the spin-spin coupling
constants. The coupling constants recorded are
$J_{12}=69.8 $ Hz, $J_{23}= -129.0$ Hz, and
$J_{13}=47.5$ Hz. Decoherence is not a major issue
in this system, with average fluorine longitudinal
$T_1$ relaxation times of $5.0$ seconds and $T_2$
relaxation times of $1.0$ seconds respectively.  The
structure of the three-qubit molecule as well as
the equilibrium NMR spectrum obtained after a
$\pi/2$ readout pulse are shown in
Fig.~\ref{system}. The resonance lines of each
qubit are labeled by the corresponding 
states of the other two coupled qubits.  
All experiments were performed 
at room temperature on a Bruker 
Avance III 400 MHz
NMR spectrometer equipped with a 
z-gradient BBO probe.
The three
fluorine nuclei cover a very large bandwidth of 68
ppm. Standard shaped pulses (of duration $400 \mu$s)  
were hence modulated to achieve uniform
excitation of all the three qubits by exciting
smaller bandwidths simultaneously at different
offsets.  Individual qubits were addressed using
low power 'Gaussian' shaped selective pulses of
$265 \mu$s duration.
Before implementing the entangling circuits,
the system was first initialized into the $\vert 000
\rangle$ pseudopure state by the spatial
averaging technique~\cite{cory-physicad}, with the
density operator given by
\begin{equation}
\rho_{000} = \frac{1-\epsilon}{8} I_8
+ \epsilon \vert 000 \rangle \langle 000 \vert
\end{equation}
with a thermal polarization $\epsilon \approx
10^{-5}$ and $I_8$ being an $8 \times 8$
identity matrix.  The experimentally created
pseudopure state $\vert 000 \rangle$ was
tomographed with a fidelity of $0.99$.
All experimentally generated states were completely
characterized by performing NMR state 
tomography~\cite{chuang-proc-98}. 
A  modified tomographic
protocol has been proposed~\cite{leskowitz-pra-04}, wherein 
a set of 7 operations defined by  
$\{ {\rm III, XXX, IIY, XYX, YII, XXY, IYY} \}$
is performed 
on the system before recording the signal.
Here $X (Y)$ 
denotes a single spin operator and $I$ is the identity operator. 
These operators can be
implemented by applying the corresponding spin selective $\pi/2$
pulses.
Motivated by this modified tomographic protocol, 
we used an expanded set of 11 operations defined by
$\{ {\rm III, IIX, IXI, XII, IIY, 
IYI, YII, YYI, IXX, XXX, YYY} \}$ 
to determine all the 63 variables for our 
system of three
qubits. 
We needed a slightly expanded set in order
to perform experimentally
accessible measurements that were sufficient
to completely characterize the experimental density
matrix with good fidelity. 
As a measure of the fidelity of the experimentally
reconstructed density matrices, we use~\cite{weinstein-prl-01}:
\begin{equation}
F =
\frac{Tr(\rho_{\rm theory}^{\dag}\rho_{\rm expt})}
{\sqrt(Tr(\rho_{\rm theory}^{\dag}\rho_{\rm theory}))
\sqrt(Tr(\rho_{\rm expt}^{\dag}\rho_{\rm expt}))}
\label{fidelity}
\end{equation}
where $\rho_{\rm theory}$ and $\rho_{\rm expt}$ denote the
theoretical and experimental density matrices
respectively.

\subsection{Generic state implementation} 
\begin{figure}[h]
\includegraphics[scale=1]{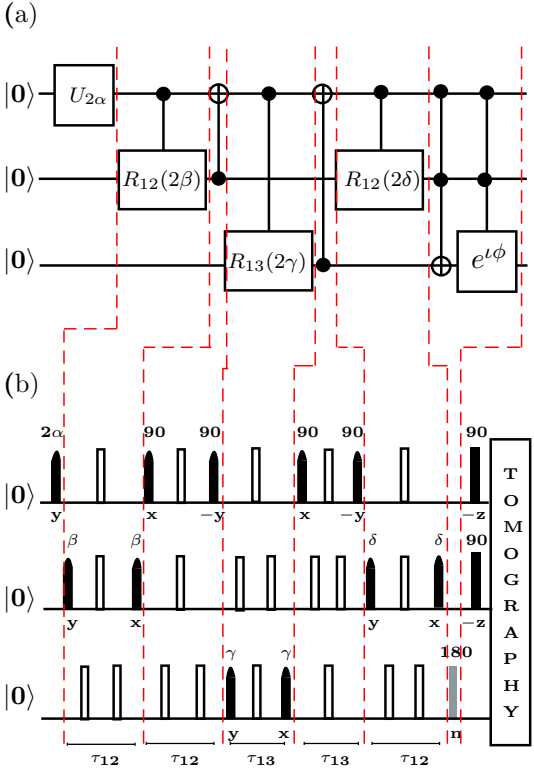}
\caption{(Color online) (a) Quantum circuit showing
the specific sequence of implementation of the 
controlled-rotation, controlled-NOT, 
controlled-controlled-NOT and
controlled-controlled-phase gates required
to construct a generic state and (b) NMR pulse sequence
to implement a general three-qubit generic state;
$\tau_{ij}$ is the evolution period under the $J_{ij}$
coupling. The
$180^{0}$ pulses are represented by unfilled
rectangles.
The other pulses are labeled with their
specific flip angles and phases. The last pulse (gray
shaded) on the third
qubit is a transition-selective $180^0$ pulse on the $\vert
110\rangle$ to $\vert 111\rangle$
transition 
about an arbitrary axis $\hat{n}$
which is inclined at angle ($\phi+90$) with the $x$-axis.
The last two rectangular pulses on the first and
second qubits are $90^{0}$ $z$-rotations, to
compensate the extra phases acquired (as 
described in the text).
\label{gen-ckt}
}
\end{figure}

The canonical (generic) state for three qubits proposed
in~\cite{acin-prl-00} is given by:
\begin{eqnarray}
& \vert \psi \rangle =  a_1 \vert 000 \rangle
+ a_2 \vert 001 \rangle + a_3 \vert 010 \rangle
+ a_4 \vert 100 \rangle + a_5 e^{i \phi} \vert 111
\rangle &
\nonumber \\
& a_i \ge 0; \quad \sum_i a_i^2=1 &  
\label{generic_state}
\end{eqnarray}
The normalization condition leads to reduction of
one parameter and hence the
state has five independent non-zero, real parameters (four
modulii and one phase). 
The state is symmetric under permutations of
the qubits and the five 
components which are invariant under local
unitaries (single-qubit operations) are the minimal number of non-local
parameters required to completely specify the
state. 
Any three-qubit state 
up to local unitaries, 
can hence be written
in the form given in
Eqn.~(\ref{generic_state}).
We base our
experimental construction on this canonical form
and  will
henceforth refer to it as the generic three-qubit
state.  
The generic three-qubit
state can be constructed 
by a sequence of gates,
starting from the system in a
pseudopure state. These gates are one-parameter unitary
transformations and as will be shown, have elegant
decompositions in terms of NMR pulses. The
normalization condition is automatically satisfied
as the normalization will be preserved under these
unitary operations.

The sequence of gates with
four real parameters $\alpha, \beta, \gamma,
\delta$ representing the amplitude parameters $a_1
\cdots a_5$ and the phase $\phi$  leading to 
the construction of a generic three-qubit state
is detailed below:
\begin{eqnarray}
&\vert 0 0 0 \rangle&  \stackrel{U^{1}_{2
\alpha}}{\longrightarrow} 
\cos{\alpha} \vert 0 0 0 \rangle +
\sin{\alpha} 
\vert 1 0 0 \rangle \nonumber \\
&\stackrel{{\rm CROT}_{12}^{2 \beta}}{\longrightarrow} &
\cos{\alpha} \vert 0 0 0 \rangle +
\sin{\alpha}
\cos{\beta}
\vert 1 0 0 \rangle  
 + \sin{\alpha} \sin{\beta} \vert 1 1 0 \rangle \nonumber \\
&\stackrel{{\rm CNOT}_{21}}{\longrightarrow} &
\cos{\alpha} \vert 0 0 0 \rangle +
\sin{\alpha}
\cos{\beta}
\vert 1 0 0 \rangle  
 + \sin{\alpha} \sin{\beta} \vert 0 1 0 \rangle \nonumber \\
&\stackrel{{\rm CROT}_{13}^{2 \gamma}}{\longrightarrow} &
\cos{\alpha} \vert 0 0 0 \rangle +
\sin{\alpha} 
\cos{\beta} \cos{\gamma}
\vert 1 0 0 \rangle  \nonumber \\
&& + \sin{\alpha} \cos{\beta} \sin{\gamma} 
\vert 1 0 1 \rangle  
+ \sin{\alpha} \sin{\beta} 
\vert 0 1 0 \rangle \nonumber \\
&\stackrel{{\rm CNOT}_{31}}{\longrightarrow} &
\cos{\alpha} \vert 0 0 0 \rangle +
\sin{\alpha}
\cos{\beta} \cos{\gamma}
\vert 1 0 0 \rangle  \nonumber \\
&& + \sin{\alpha} \cos{\beta} \sin{\gamma} 
\vert 0 0 1 \rangle  
+ \sin{\alpha} \sin{\beta} 
\vert 0 1 0 \rangle \nonumber \\
&\stackrel{{\rm CROT}_{12}^{2 \delta}}{\longrightarrow} &
\cos{\alpha} \vert 0 0 0 \rangle +
\sin{\alpha}
\cos{\beta}
\cos{\gamma} \cos{\delta} \vert 1 0 0 \rangle  \nonumber \\ 
&& + \sin{\alpha} \cos{\beta} \cos{\gamma} \sin{\delta}
\vert 1 1 0 \rangle \nonumber \\
&& + \sin{\alpha} \cos{\beta}
\sin{\gamma} \vert 0 0 1 \rangle
 + \sin{\alpha} \sin{\beta} 
\vert 0 1 0 \rangle   \nonumber \\
&\stackrel{{\rm CCN}_{12,3}}{\longrightarrow} &
\cos{\alpha} \vert 0 0 0 \rangle +
\sin{\alpha}
\cos{\beta}
\cos{\gamma} \cos{\delta} \vert 1 0 0 \rangle  \nonumber \\ 
&& + \sin{\alpha} \cos{\beta} \cos{\gamma} \sin{\delta}
\vert 1 1 1 \rangle \nonumber \\
&& + \sin{\alpha} \cos{\beta}
\sin{\gamma} \vert 0 0 1 \rangle
 + \sin{\alpha} \sin{\beta} 
\vert 0 1 0 \rangle   \nonumber \\
&\stackrel{{\rm Ph}_{12,3}^{\phi}}{\longrightarrow} &
\cos{\alpha} \vert 0 0 0 \rangle +
\sin{\alpha} \cos{\beta}
\sin{\gamma} \vert 0 0 1 \rangle \nonumber \\
&& + \sin{\alpha} \sin{\beta} 
\vert 0 1 0 \rangle   
+ \sin{\alpha}
\cos{\beta}
\cos{\gamma} \cos{\delta} \vert 1 0 0 \rangle  \nonumber \\ 
&& + e^{\iota \phi} \sin{\alpha} \cos{\beta} \cos{\gamma}
\sin{\delta}
\vert 1 1 1 \rangle \nonumber \\
\label{genstate-nmr}
\end{eqnarray}
The operator $U_{2 \alpha}^{1}$ 
is a separable, non-entangling transformation
belonging to the $SU(2)$ group which implements  
a rotation by an arbitrary angle $\alpha$ on
the first qubit, 
leading to a generalized superposition
state of the qubit.
The global phase is not detectable in NMR experiments and is thus
ignored throughout in gate implementation;
CROT$_{ij}^{2 \theta}$ implements a controlled
rotation by an arbitrary angle $\theta$, with the
$i^{th}$ qubit as control and $j^{th}$ as target;
CNOT$_{ij}$ implements a controlled-NOT gate, with the
$i^{th}$ qubit as control and $j^{th}$ as target;
CCN$_{12,3}$ 
implements a controlled-controlled-NOT (Toffoli) gate 
on the $3$rd qubit i.e.
it flips the state of qubit $3$, if and only if 
both qubits $1$ and $2$ are in
the $\vert 1 \rangle$ state;
Ph$_{12,3}^{\phi}$ is a
controlled-controlled-phase shift gate with $1,2$
as control qubits and $3$ being the target qubit.
The state thus obtained has five variables:
$\alpha \in [0, \pi/2], \beta \in [0, \pi/2],
\gamma \in [0, \pi/2], \delta \in [0, \pi/2]$
and $\phi \in [0, 2\pi]$.

The quantum circuit for generic state construction
is given in Fig.~\ref{gen-ckt}(a).  The circuit
consists of a single-qubit rotation gate, followed
by several two-qubit controlled-rotation and
controlled-NOT gates, a three-qubit
controlled-controlled NOT (Toffoli) gate, and
finally a controlled-controlled phase gate that
introduces a relative phase in the $\vert 1 1 1
\rangle$ state.

The NMR pulse sequence to construct the generic
three-qubit state starting from the pseudopure
state $\vert 0 0 0 \rangle$ is given in
Fig.~\ref{gen-ckt}(b).  Refocusing pulses are used
in the middle of all J-evolution periods to
compensate for chemical shift evolution.  Pairs of
$\pi$ pulses have been inserted at 1/4 and 3/4 of
the J-evolution intervals to eliminate undesirable
evolution due to other J-couplings.  The $180^{0}$
pulses are represented by unfilled rectangles,
while the other pulses are labeled with their
specific flip angles and phases. 
An ideal
controlled rotation gate CROT$_{ij}$, where `$i$'
is control and `$j$' is the target qubit ($i < j$)
is implemented by the sequence~: $(\theta)_{-y}^j
\, (\frac{\pi}{2})_{z}^{i,j} \, \frac{1}{4J_{ij}}
\, (\pi)_{y}^{i,j} \frac{1}{4J_{ij}} \,
(\pi)_{y}^{i,j} \, (\theta)_{-y}^j \,
(\pi)_{z}^{i,j}$~\cite{jones-jmr-1998}; here
$(\theta)_{\alpha}^{i}$ denotes an rf pulse of
flip angle $\theta$ and phase $\alpha$ applied on
the $i$th qubit, $(\beta)_{\alpha}^{i,j}$ denotes
an rf pulse of flip angle $\beta$ and phase
$\alpha$ applied simultaneously on both the $i$th
and $j$th qubits, and $\frac{1}{4 J_{ij}}$ denotes
an evolution period under the coupling Hamiltonian
(using standard NMR notation).  The above sequence
for the ideal CROT$_{ij}$ gate contains two
$z$-rotations on each of the control and target
qubits, which are of long duration and give rise
to experimental imperfections. In order to shorten
the gate duration and hence reduce experimental
artifacts, we implemented a shorter pulse sequence
corresponding to $(\theta)_{-y}^j \,
\frac{1}{4J_{ij}} \, (\pi)_{y}^{i,j}
\frac{1}{4J_{i,j}} \, (\pi)_{y}^{i,j} \,
(\theta)_{-x}^j $, which creates the desired state
alongwith a relative phase.  We keep track of the
relative phase gained at the end of each
controlled operation and implement $z$-rotations
on the spins at the end of the sequence to
compensate for the relative phases acquired.
The last two gates in the circuit, 
namely the controlled-controlled NOT (Toffoli) gate
and the controlled-controlled phase gate were
simultaneously implemented using a single transition-selective
$\pi$ pulse, applied about an arbitrary axis of rotation
$\hat{n}$ (gray-shaded in Fig.~\ref{gen-ckt}(b))
~\cite{dorai-jmr-1995,peng-cpl-2001}.  A
three-qubit controlled-controlled NOT (Toffoli)
gate can be experimentally realized by a transition-selective
$(\pi)_y$ pulse between energy levels $\vert 110
\rangle$ and $\vert 111 \rangle$.  
A transition-selective pulse $(\pi)_{\hat{n}}$
about an arbitrary axis of rotation $\hat{n}=\cos
\phi^{'} \hat{x} + \sin \phi^{'} \hat{y}$, on the
other hand, introduces an extra phase of $e^{\iota
\phi}$ ($\phi^{'}=\phi+\pi/2$).  Hence,
$(\pi)_{\hat{n}}^{\vert 110 \rangle \rightarrow
\vert 111 \rangle}$ when applied on the basis
vector $\vert 110 \rangle$, results in the state
$e^{\iota \phi} \vert 111 \rangle$.  This is an
ingenious method to reduce the experimental time,
and comes in handy in completing the circuit
implementation before the decoherence begins to
introduce significant distortions.

\begin{figure}[h]
\includegraphics[scale=1.0]{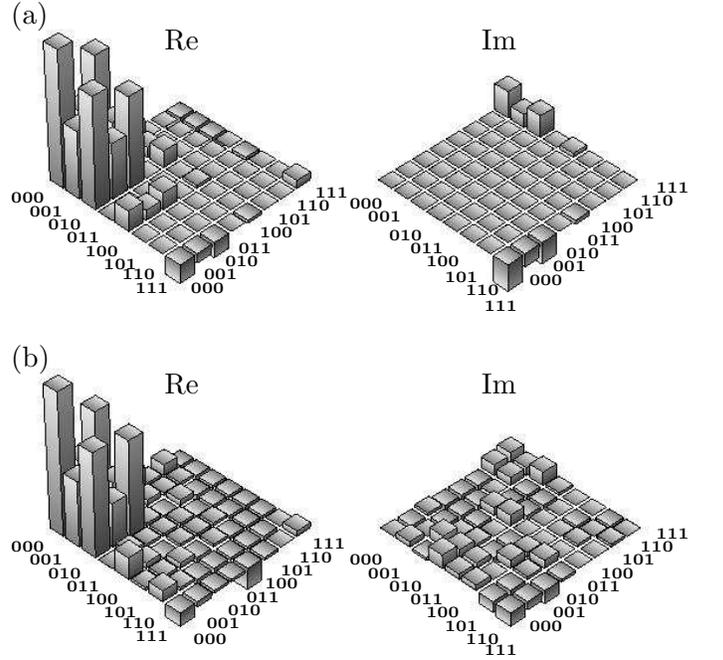}
\caption{The real (Re) and imaginary (Im) parts of the
(a) theoretical and (b) experimental 
density matrices  for the three-qubit generic state,
reconstructed using full state tomography.
The values of the parameters are $\alpha=45^{0}, 
\beta=55^{0}, \gamma=60^{0}, \delta=58^{0}, \phi=125^{0}$.
The rows and columns encode the
computational basis in binary order, from
$\vert 000 \rangle$ to $\vert 111 \rangle$.
The experimentally tomographed state has
a fidelity of $0.92$.
\label{gentomo}  
}
\end{figure}

To demonstrate our general method to create generic 
three-qubit states, we implement our scheme to 
create a state with a nontrivial structure.
We chose a state  in which all the terms in the generic 
state expression given in Eqn.~\ref{genstate-nmr} are 
involved in a nontrivial way. We have chosen 
$\alpha=45^0,
\beta= 55^0,
\gamma= 60^0,
\delta= 58^0$ and $\phi=125^0$.
This set of parameters leads to the creation of
the generic state:
\begin{eqnarray}
&& 0.707 \vert 000 \rangle +
0.351 \vert 001 \rangle +
0.579 \vert 010 \rangle +
0.107 \vert 100 \rangle + \nonumber \\
&& 0.172 e^{i (125^0)} \vert 111 \rangle
\end{eqnarray}
The tomograph
corresponding to this state is shown in
Fig.~\ref{gentomo}, wherein the
experimentally tomographed state 
(Fig.~\ref{gentomo}(b))
is compared with the theoretically
expected state (Fig.~\ref{gentomo}(a)). The fidelity of the 
experimentally
tomographed state (by the definition given in
Eqn.~\ref{fidelity}) in this case is $0.92$. 

Our method is quite general and can be used to construct any 
generic state of the three-qubit system. Given that the
relaxation times for our system are quite long and the qubits are well
separated in frequency space, it is also possible to 
perform single-qubit operations to transform the state
further. 
\subsection{GHZ state implementation}
\begin{figure}[h]
\includegraphics[scale=1]{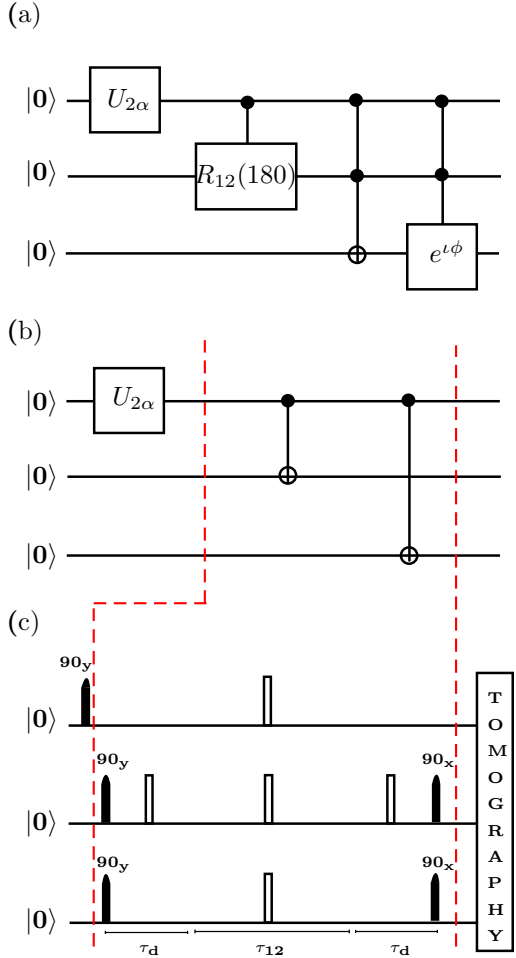}
\caption{(Color online) (a) Quantum circuit to implement a
generalized GHZ state, derived from the
general circuit for generic state
construction given in Fig.~\ref{gen-ckt}(a).
(b) Simplified circuit for experimental
implementation of the GHZ state.
(c) NMR pulse sequence corresponding to
the circuit in (b).
The
$\tau_d = \frac{\tau_{13} - \tau_{12}}{2}$ period is
tailored such that the
system evolves solely under the $J_{13}$ 
coupling term.
\label{ghz-ckt}
}
\end{figure}

Generalized GHZ states are a special case of the
generic state given in Eqn.~\ref{generic_state},
corresponding to the parameter values
$\alpha= \alpha, \beta= \gamma=0, \delta= \pi/2, \phi=0$,  
and the circuit given in
Fig.~\ref{gen-ckt}(a) reduces to 
the circuit given in Fig.~\ref{ghz-ckt}(a).
The two controlled-rotation gates
CROT$_{12}^{2\beta}$ and CROT$_{13}^{2\gamma}$
are hence redundant for the state
implementation and the 
simplified experimental circuit
is given in Fig.~\ref{ghz-ckt}(b),
with
a single-qubit
rotation followed by two controlled-NOT gates.
An arbitrarily
weighted GHZ kind of entangled state can be
prepared from the initial pseudopure state $\vert
0 0 0 \rangle$ by the sequence of operations
\begin{eqnarray}
\vert 0 0 0 \rangle &\stackrel{U_{2\alpha}^1}{\longrightarrow}&
\cos{\alpha} \vert 0 0 0 \rangle + 
\sin{\alpha} \vert 1 0 0 \rangle \nonumber \\
&\stackrel{\rm CNOT_{12}}{\longrightarrow}&
\cos{\alpha} \vert 0 0 0 \rangle + 
\sin{\alpha} \vert 1 1 0 \rangle \nonumber \\
&\stackrel{\rm CNOT_{13}}{\longrightarrow}&
\cos{\alpha} \vert 0 0 0 \rangle + 
\sin{\alpha} \vert 1 1 1 \rangle 
\end{eqnarray}
For $\alpha = \pi/4$, the above sequence leads to a pure
GHZ 
state~\cite{laflamme-proc-98,nelson-pra-00,teklemariam-pra-02}:
\begin{equation}
\vert \psi_{\rm GHZ} \rangle = \frac{1}{\sqrt{2}}(\vert 000 \rangle
+ \vert 111 \rangle)
\end{equation}
\begin{figure}[h]
\includegraphics[scale=1.0]{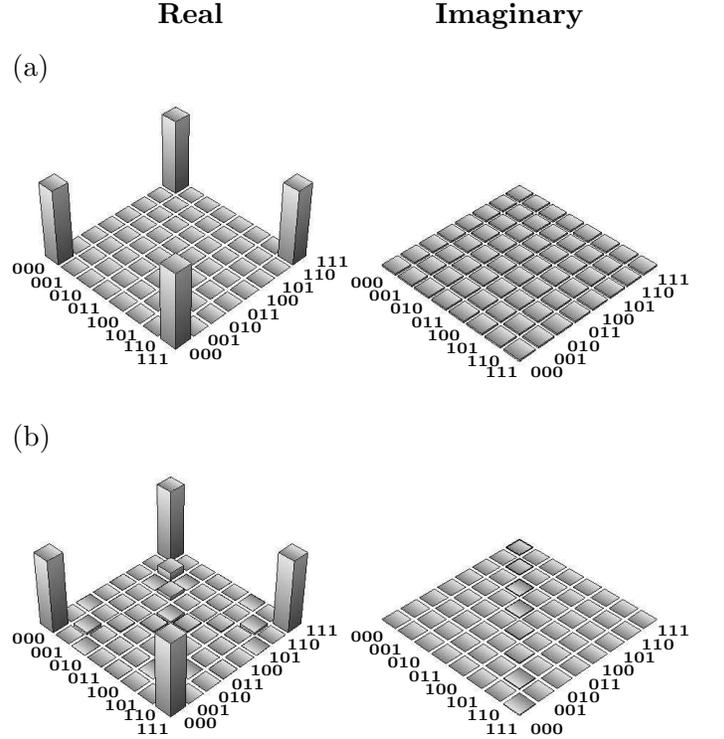}
\caption{The real (Re) and imaginary (Im) parts of the
(a) theoretical and (b) experimental 
density matrices  for the GHZ state,
reconstructed using full state tomography.
The rows and columns encode the
computational basis in binary order, from
$\vert 000 \rangle$ to $\vert 111 \rangle$.
The experimentally tomographed state has
a fidelity of 0.97.
\label{ghztomo}
}
\end{figure}

The quantum circuit and the NMR pulse sequence
used to create an arbitrary GHZ-like entangled
state beginning from the pseudopure state $\vert 0
0 0 \rangle$ and ignoring overall phase factors
are given in Fig.~\ref{ghz-ckt}(b) and (c) respectively.  
The CNOT$_{12}$ and CNOT$_{13}$
in the circuit are controlled-NOT
gates with qubit 1 as the control and qubit 2 (3)
as the target. Since the target qubits are
different in both these cases, these gates commute
and can be applied in parallel, leading to
a reduction in experimental time. For our system
$\tau_{13} > \tau_{12}$, where $\tau_{ij}$ denotes
the evolution period  under 
the $\frac{1}{2J_{ij}}$ coupling term.
Hence, during the period
$\tau_{12}$, both qubits 2 and 3 evolve under the
the J-couplings $J_{12}$ and $J_{13}$
(Fig.~\ref{ghz-ckt}(c)).
The evolution in the intervals
$\displaystyle \tau_d = \frac{\tau_{13}-\tau_{12}}{2}$ is solely governed by
the $J_{13}$ coupling term, and by the end of 
the evolution period, the system evolves under $J_{12}$
and $J_{13}$ couplings for durations
$\frac{1}{2J_{12}}$ and
$\frac{1}{2J_{13}}$ respectively.
The state
generated experimentally (Fig.~\ref{ghztomo}(b))
was tomographed and lies very close to the
theoretically expected state
(Fig.~\ref{ghztomo}(a)) with a computed fidelity
of 0.97. 
\subsection{W-state implementation} 
\begin{figure}[h]
\includegraphics[scale=1]{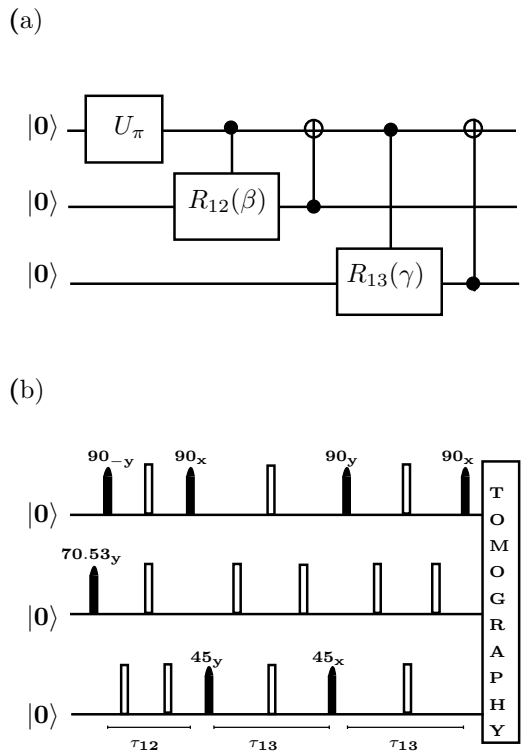}
\caption{(Color online) (a) Quantum circuit to implement the
W-state, derived from the general circuit for
generic state construction given in 
Fig.~\ref{gen-ckt}(a). (b) Simplified
circuit for experimental implementation
of the W-state.
(c) NMR pulse sequence
to experimentally implement the W-state,
starting from the initial pseudopure 
state $\vert 1 0 0 \rangle$. The first pulse
on the second qubit implements a 
$U^{2}_{2 \beta}$ rotation, with
$2 \beta = 2 \sin^{-1}{(1/\sqrt{3})} 
\equiv 70.53^{0}$.
\label{wstate-ckt}
}
\end{figure}

Generalized W-states are another special case of
the generic state given in
Eqn.~\ref{generic_state}, corresponding to the
parameter values $ \alpha=\pi/2, \beta, \gamma \in
[0, \pi/2], \delta=0, \phi=0$, leading to the
state $\vert \psi \rangle = \cos{\gamma}
\cos{\beta} \vert 1 0 0 \rangle + \sin{\gamma}
\cos{\beta} \vert 0 0 1 \rangle + \sin{\beta}
\vert 0 1 0 \rangle$.  The circuit for generalized
W-states derived from the circuit in
Fig.~\ref{gen-ckt}(a) is given in
Fig.~\ref{wstate-ckt}(a) and can be constructed
by the
sequential operation of the gates:

\begin{eqnarray}
\vert 000 \rangle & \stackrel{U^{1}_{\pi}}{\longrightarrow} &
\vert 100 \rangle \nonumber \\
 & \stackrel{\rm CROT_{12}^{2\beta}} {\longrightarrow} &   \cos \beta
\vert 100 \rangle + \sin \beta \vert 110 \rangle
\nonumber \\
& \stackrel{\rm CNOT_{21}} {\longrightarrow} &  \cos \beta \vert 100
\rangle +
\sin \beta \vert 010 \rangle
\nonumber \\
& \stackrel{\rm CROT_{13}^{2\gamma}}{\longrightarrow} &  \cos \gamma
\cos \beta \vert 100 \rangle + \sin \gamma \cos \beta
\vert 101 \rangle + \sin \beta \vert 010 \rangle 
\nonumber \\
& \stackrel{\rm CNOT_{31}} {\longrightarrow} &  \cos \gamma \cos \beta
\vert 100 \rangle + \sin \gamma \cos \beta \vert
001 \rangle +
\sin \beta \vert 010 \rangle \nonumber \\
\label{weqn}
\end{eqnarray}

\begin{figure}[h]
\includegraphics[scale=1.0]{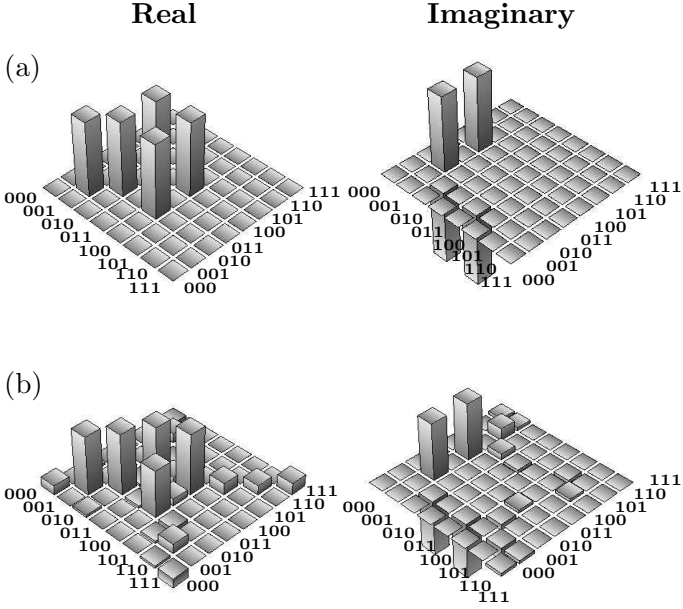}
\caption{The real (Re) and imaginary (Im) parts of the
(a) theoretical and (b) experimental 
density matrices  for the W state,
reconstructed using full state tomography.
The rows and columns encode the
computational basis in binary order, from
$\vert 000 \rangle$ to $\vert 111 \rangle$.
The experimentally tomographed state has a fidelity
of 0.96.
\label{wtomo}
}
\end{figure}

The first gate in the circuit, namely a rotation
by $\pi$ on the first qubit, can be avoided
by starting the implementation on a different
initial state.  We hence begin with the pseudopure
state $\vert 1 0 0 \rangle$ as the initial state
in our experiments. We also avoid implementing the
second gate in the circuit in Eqn.(~\ref{weqn}),
namely the controlled-rotation ${\rm
CROT}_{12}^{2\beta}$ gate, and instead implement the
much simpler $U^{2}_{2\beta}$ gate on the second
qubit, which in this case yields the same result.
For $2 \beta=
2 \sin^{-1}{(1/\sqrt{3})}$ and $\gamma=45^{0}$, the
circuit leads to implementation of the
standard W-state upto a phase factor
\begin{equation}
\vert \psi_{\rm W}  \rangle = \frac{1}{\sqrt{3}}(i \vert 001 \rangle
+ \vert 010 \rangle + \vert 100 \rangle) 
\end{equation}
One can get rid of the extra phase factor by a
single-qubit unitary gate.
The simplified experimental
circuit and the NMR pulse sequence for the creation of an
arbitrary W-like entangled state beginning from
the pseudopure state $\vert 1 0 0 \rangle$ and
ignoring overall phase factors, are given in
Figs.~\ref{wstate-ckt}(b) and (c) respectively.  The experimentally
reconstructed density matrix (Fig.~\ref{wtomo}(b))
matches well with the theoretically expected
values (Fig.~\ref{wtomo}(a)), with a computed
state fidelity of 0.96. 
\section{Three-qubit state reconstruction from
two-party reduced states}
\label{2tomo}
Linden et al. discovered a surprising fact
about multiparty correlations, namely, that 
``the parts determine the whole for a generic
pure state''~\cite{linden-prl-1-02,linden-prl-2-02}. 
For three qubits, this
implies that all the information in a generic
three-party state is contained in its three
two-party reduced states, which then uniquely
determine the full three-party state. 
The only
exceptions to the above hypothesis are the
generalized GHZ states, and no set of their
reduced states can uniquely determine such
entangled states.  
This is an important result which sheds some
light on how information is stored in
multipartite entangled states. In a related
work, Diosi et al.~\cite{diosi-pra-04}
presented a tomographic protocol to completely
characterize almost all generic three-qubit
pure states, based only on pairwise two-qubit
detectors.
\begin{figure}[h]
\includegraphics[scale=1.0]{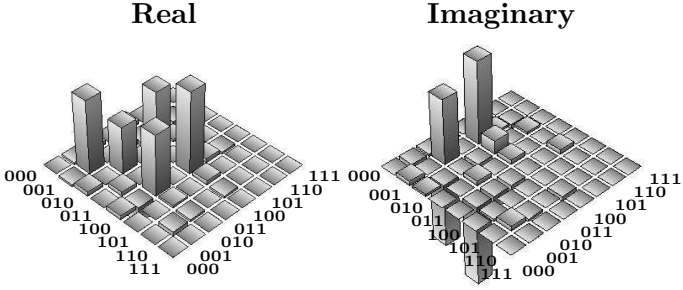}
\caption{The real (Re) and imaginary (Im) parts 
of the density matrix for the W-state: 
(a) The two-qubit reduced density matrix 
$\rho_{AB}$. (b) The two-qubit reduced
density matrix $\rho_{BC}$.
(c) The entire
three-qubit density matrix 
$\rho_{ABC}$,
reconstructed from the
corresponding two-qubit reduced density
matrices.
The rows and columns encode the
computational basis in binary order, from
$\vert 00 \rangle$ to $\vert 11 \rangle$ for
two qubits and from 
$\vert 000 \rangle$ 
to $\vert 111 \rangle$ for three qubits.
The tomographed state has a fidelity
of 0.97.
\label{w-2bit}
}
\end{figure}

In this paper we describe the first experimental
demonstration of this interesting quantum
mechanical feature of three-qubit states.
We use the same algorithm delineated by
Diosi et al.~\cite{diosi-pra-04}, to reconstruct
three-qubit states from their two-party
reduced states.
Let us consider a three-qubit pure state
$\rho_{ABC}=\ket{\psi_{ABC}}\bra{\psi_{ABC}}$,
with $\rho_{AB}$, $\rho_{BC}$, $\rho_{AC}$ being
its two-party reduced states.  The single-qubit
reduced states $\rho_{A}$, $\rho_{B}$ and
$\rho_{C}$ can be further obtained from the
two-party reduced states.  Since $\rho_{ABC}$ is
pure, $\rho_{A}$ and $\rho_{BC}$ share the same
set of eigen values, and  can be written as
\begin{eqnarray}
\rho_A &=&\sum_i
p_A^i \ket{i} \bra{i}  \nonumber \\ 
\rho_{BC} &=& \sum_i p_A^i \ket{i;BC} \bra{i;BC}
\end{eqnarray}
where $\{\ket{i}\}$ are the eigenvectors of
$\rho_{A}$ with eigenvalues $\{p_A^i\}$, 
and $\{ \ket{i;BC}\}$ are the eigenvectors of
$\rho_{BC}$ with eigenvalues $\{ p_A^i\}$.  
The
three-qubit states compatible with $\rho_{A}$ and
$\rho_{BC}$ are 
\begin{equation}
\ket{\psi_{ABC};\alpha}=\sum_i
e^{\iota \alpha_i}
\sqrt{p_A^i}\ket{i}\otimes\ket{i;BC}
\end{equation}
Using a
similar argument, 
the set
of three-qubit pure states obtained from
$\rho_{AB}$ and $\rho_C$  is given by
\begin{equation}
\ket{\psi_{ABC};\gamma}=\sum_k
e^{\iota \gamma_k}
\sqrt{p_c^k}\ket{k;AB}\otimes\ket{k} 
\end{equation}
where $\{\ket{k}\}$ are
the eigenvectors of $\rho_C$ with eigenvalues
$\{p_c^k\}$ and $\{\ket{k;AB}\}$ are the
corresponding eigenvectors of $\rho_{AB}$.  Since
the pure state $\ket{\psi_{ABC}}$ is compatible
with both $\rho_{AB}$ and $\rho_{BC}$, we can
determine the values of $\alpha_i$ and $\gamma_k$
such that
$\ket{\psi_{ABC};\alpha}=\ket{\psi_{ABC};\gamma}$.
We thus obtain almost all three-qubit
pure states from any two of their corresponding
two-party reduced states. 
The set ($\rho_{AB}$, $\rho_{AC}$)
or the equivalent set ($\rho_{AB}$, $\rho_{BC}$) 
can be used to reconstruct $\rho_{ABC}$.
\begin{figure}[h]
\includegraphics[scale=1.0]{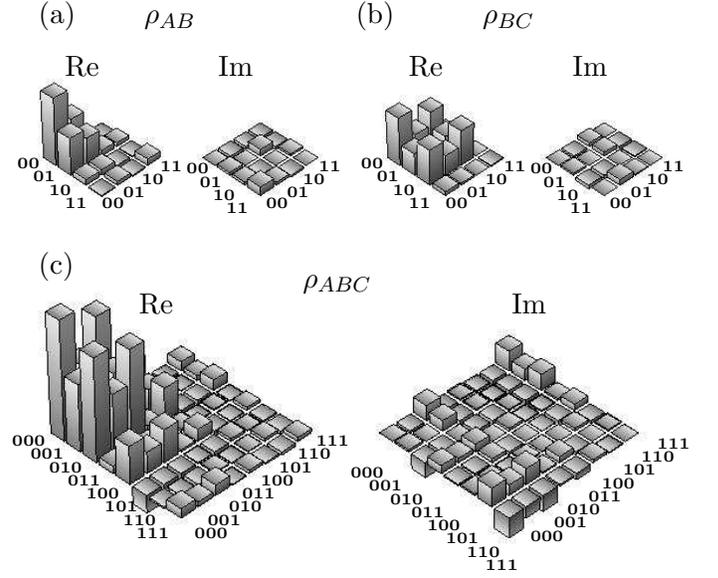}
\caption{The real (Re) and imaginary (Im) parts of the
density matrix for the generic state:
(a) The two-qubit reduced density matrix $\rho_{AB}$.
(b) The two-qubit reduced density matrix $\rho_{BC}$.
(c) The entire
three-qubit density matrix $\rho_{ABC}$, 
reconstructed from the 
corresponding two-qubit reduced density
matrices. The parameter set includes
$\alpha=45^{0}, \beta=55^{0}, \gamma=60^{0},
\delta=58^{0}, \phi=125^{0}$.
The rows and columns encode the
computational basis in binary order, from
$\vert 00 \rangle$ to $\vert 11\rangle$ for
two qubits and from
$\vert 000 \rangle$ 
to $\vert 111 \rangle$ for three qubits.
The tomographed state has a fidelity
of 0.90.
\label{gen-2bit}
}
\end{figure}

The two-party reduced
states $\rho_{AB}$, $\rho_{BC}$ and
$\rho_{AC}$ were computed by performing partial state
tomography.  
The set of tomography operations performed to
experimentally reconstruct all three two-party
reduced states include:~\{III, IXI, IYI, XXI\} to
reconstruct $\rho_{AB}$; \{III, IIX, IIY, IXX\} to
reconstruct $\rho_{BC}$ and \{III, IIX, IIY, XIX\} to
reconstruct $\rho_{AC}$.  Almost any three-qubit
pure state $\rho_{ABC}$ (except those belonging to the
generalized GHZ class) can be determined by
choosing any two sets from the above. 
The three-party state $\rho_{ABC}$ reconstructed using 
the $(\rho_{AB}, \rho_{BC})$
set of two-party reduced states was compared with
the same state reconstructed using complete
tomography, and the results match well.
For the W state
we tomographed $\rho_{AB}$ and $\rho_{BC}$ to give us
\begin{eqnarray}
 \rho_{AB}&=&
\left(
\begin{array}{cccc}
 0.36 & 0. & 0. & 0.\, -0.01 i \\
 0. & 0.21 & 0.2\, +0.05 i & -0.01 \\
 0. & 0.2\, -0.05 i & 0.21 & 0.01 \\
 0.\, +0.01 i & -0.01 & 0.01 & 0.22 \\
\end{array}
\right)
\nonumber \\
 \rho_{BC}&=&
\left(
\begin{array}{cccc}
 0.34 & -0.01 & 0.\, +0.01 i & 0 \\
 -0.01 & 0.3 & 0.\, +0.24 i & 0.02 \\
 -0.01 & 0.\, -0.24 i & 0.2 & 0. \\
 0 & 0.02 & 0 & 0.16 \\
\end{array}
\right)
\end{eqnarray}
These experimental tomographed density matrices were then
used to reconstruct 
the three-qubit W-state density matrix
$\rho_{ABC}$. The thus reconstructed $\rho_{ABC}$ is given by
\begin{widetext}
\begin{equation}
\rho_{ABC}=
 \left(
\begin{array}{cccccccc}
 0. & -0.01-0.02 i & 0.02\, -0.01 i & 0. & 0.02 & 0. & 0. & 0. \\
 -0.01+0.02 i & 0.36 & 0.\, +0.29 i & 0.02 & -0.1+0.37 i & -0.01+0.01 i & -0.02-0.02 i & 0. \\
 0.02\, +0.01 i & 0.\, -0.29 i & 0.23 & 0.\, -0.02 i & 0.3\, +0.08 i & 0.01 & -0.01+0.01 i & 0. \\
 0. & 0.02 & 0.\, +0.02 i & 0. & -0.01+0.02 i & 0. & 0. & 0. \\
 0.02 & -0.1-0.37 i & 0.3\, -0.08 i & -0.01-0.02 i & 0.4 & 0.02 & -0.01+0.02 i & 0. \\
 0. & -0.01-0.01 i & 0.01 & 0. & 0.02 & 0. & 0. & 0. \\
 0. & -0.02+0.02 i & -0.01-0.01 i & 0. & -0.01-0.02 i & 0. & 0. & 0. \\
 0. & 0. & 0. & 0. & 0. & 0. & 0. & 0. \\
\end{array}
\right)
\end{equation}
\end{widetext}
The reconstructed density matrix for the W-state
is shown in Fig.~\ref{w-2bit}, computed from two sets
of the corresponding two-qubit reduced density
matrices. The tomographed state has a fidelity of
0.97, which matches well with the fidelity of the
original three-qubit density matrix of the W-state
(Fig.~\ref{wtomo}(b)).
As another illustration of reconstructing the
whole state from its parts, the reconstructed
density matrix of the experimentally generated
generic state
with a parameter set: 
$\alpha=45^{0}, \beta=55^{0}, 
\gamma=60^{0}, \delta=58^{0},
\phi=125^{0}$,
is shown in Fig.~\ref{gen-2bit}. The two-party
reduced states were able to reconstruct this
three-qubit state with a fidelity of 0.90, which
compares well with the full reconstruction of the
entire three-qubit state given in 
Fig.~\ref{gentomo}(b). 
\section{Concluding Remarks}
\label{concl}
We have proposed and implemented an NMR-based scheme to
construct a generic three-qubit state from which any general
pure state of three-qubits (including separable, biseparable
and maximally entangled states) can be constructed, up to
local unitaries.  Full tomographic reconstruction of the
experimentally generated states showed good fidelity of
preparation and we have achieved a high degree of control
over the state space of three-qubit quantum systems.
Generating generic three-qubit states with a nontrivial
phase parameter was an experimental challenge and we archived
it by crafting a special pulse scheme.  It has been
previously shown that in a system of three qubits, no
irreducible three-party correlations exist and that all
information about the full quantum state is completely
contained in the three two-party correlations. We have
demonstrated this important result experimentally in a
system of three qubits.  The three-qubit density operator
$\rho_{ABC}$ is obtained by complete quantum state
tomography and compared with the same three-qubit state
reconstructed from tomographs of the two-party reduced
density operators given by $\rho_{AB}$, $\rho_{BC}$ and
$\rho_{AC}$.  It is expected that our experiments will pave
the way for an understanding of how information is stored in
multi-partite entangled systems.
\begin{acknowledgments}
All experiments were performed on a Bruker Avance-III
400 MHz FT-NMR spectrometer at the NMR Research Facility
at IISER Mohali. SD acknowledges UGC
India for financial support.
\end{acknowledgments}
%
\end{document}